\documentclass[twocolumn,aps,pra,linenumber,showpacs,floatfix,superscriptaddress]{revtex4-1}
\usepackage{amsfonts}
\usepackage{amsmath}
\usepackage{amssymb}
\usepackage{dcolumn}
\usepackage{graphicx}%
\usepackage{bm}
\usepackage{color}

\begin{document}

\newcommand{\HIM}{
Helmholtz Institute Mainz, D-55128 Mainz, Germany}

\newcommand{\NZIAS}{
Centre for Theoretical Chemistry and Physics,
The New Zealand Institute for Advanced Study,
Massey University Auckland, Private Bag 102904, 0745 Auckland, New Zealand}

\newcommand{\UNSW}{
School of Physics, University of New South Wales, Sydney 2052, Australia}

\newcommand{\MBU}{
Department of Chemistry, Faculty of Natural Sciences, Matej Bel University, 
Tajovsk\'{e}ho 40, SK-974 00 Bansk\'{a} Bystrica, Slovakia}

\newcommand{\PUM}{Fachbereich Chemie, Philipps-Universit\"{a}t Marburg, 
Hans-Meerwein-Str., D-35032 Marburg, Germany}

\title{Nuclear-spin dependent parity violation in diatomic molecular ions}


\author{A. Borschevsky}
\affiliation{\NZIAS}
\affiliation{\HIM}

\author{M. Ilia\v{s}}
\affiliation{\MBU}

\author{V. A. Dzuba}
\affiliation{\UNSW}

\author{K. Beloy}
\altaffiliation{Present address: National Institute of Standards and Technology,
Boulder, Colorado 80305}
\affiliation{\NZIAS}

\author{V. V. Flambaum}
\affiliation{\UNSW}
\affiliation{\NZIAS}

\author{P. Schwerdtfeger}
\affiliation{\NZIAS}
\affiliation{\PUM}

\date{\today}


\begin{abstract}
Nuclear-spin-dependent (NSD) parity violating (PV) effects can be strongly enhanced in diatomic molecules containing heavy atoms. 
Future measurements are anticipated to provide nuclear anapole moments and strength constants for PV nuclear forces. 
In light molecules, the NSD electroweak electron-nucleus interaction may also be detected. 
Here we calculate NSD PV effects for molecular ions. Our calculations are motivated by rapid developments in 
trapping techniques for such systems at low temperatures. 

\end{abstract}

\maketitle

It was previously shown that
nuclear spin-dependent (NSD) parity violation (PV) effects are enhanced by a
factor of $10^5$ in diatomic molecules with $^2\Sigma_{1/2}$ and
$^2\Pi_{1/2}$ electronic states due to the mixing of close rotational
states of opposite parity ($\Omega$-doublet for
$^2\Pi_{1/2}$) \cite{Lab78,SusFla78,FlaKri85_2}. 
DeMille and co-workers suggested measuring
NSD PV effects by using neutral diatomic molecules in a  
Stark interference experiment to determine the mixing between
opposite-parity rotational/hyperfine levels \cite{DemCahMur08}.
Another proposal was published in Ref. \cite{ IsaHoeBer10}, and corresponding experiments have already started.  
Recently, it was demonstrated that positive molecular ions may be easily trapped and studied 
at low temperatures \cite{Odom11}, which motivated us to perform calculations on NSD PV effects in such systems.

The term in the Hamiltonian operator arising from the NSD parity violating electron-nucleus interaction is 
\begin{eqnarray}
H_A=\kappa_{NSD} \frac{G_F}{\sqrt{2}}\frac{\bm{\alpha}\cdot \mathbf{I}}{I}\rho(\mathbf{r}).
\label{Ha}
\end{eqnarray}
Here, and throughout the text, we use atomic units. In Eq.~(\ref{Ha}), $\kappa_{NSD}$ is the  dimensionless strength constant, $G_{F}=2.22249\times10^{-14}$ a.u. is the Fermi
constant, $\bm{\alpha}$ is a vector comprised of the conventional Dirac matrices, $\mathbf{I}$ is the
nuclear spin, $\mathbf{r}$ is the displacement of the valence electron from
the nucleus, and $\rho(\mathbf{r})$ is the (normalized) nuclear density. There are three sources for this interaction:
the first contribution arises from the electroweak neutral coupling between electron vector and 
nucleon axial-vector currents ($\mathbf{V}_e\mathbf{A}_N$) \cite{NovSusFla77}. 
The second contribution comes from the nuclear-spin-independent weak interaction combined with the hyperfine 
interaction \cite{FlaKri85}. Finally, the nuclear anapole moment contribution, which scales with the number of nucleons $A$
as $\kappa_{A}\sim A^{2/3}$, becomes the dominant contribution in spin-dependent atomic PV effects for a
sufficiently large nuclear charge $Z$ \cite{FlaKri80,FlaKriSus84}.

The anapole moment was first predicted by Zeldovich \cite{Zel58} in 1958 as a new parity ($P$) violating and time ($T$) reversal 
conserving moment of an elementary particle. It  appears in the
second-order multipole expansion of the magnetic vector-potential
simultaneously  
with the $P$-- and $T$-- violating  magnetic quadrupole moment
\cite{SusFlaKri84}. The nuclear anapole moment was experimentally
discovered in the $^{133}$Cs atom in 1997 \cite{WooBenCho97} following a proposal by Flambaum and
Khriplovich \cite{FlaKri80}, who showed that the nuclear anapole provides the dominant contribution to the 
nuclear-spin-dependent parity violating effect in atoms and molecules.

The nuclear anapole requires nuclear spin $I\neq 0$ and in a simple  valence  model  has the following value \cite{FlaKriSus84},
\begin{eqnarray}
\kappa_A= 1.15\times 10^{-3} \left(\frac{\mathcal{K}}{I+1} \right) A^{2/3} \mu_i g_i.
\label{eqanapole}
\end{eqnarray}
Here, $\mathcal{K}=(-1)^{I+\frac{1}{2}-l}(I+1/2)$, $l$ is the orbital
angular momentum of the external unpaired nucleon 
$i=n,p$;
 $\mu_p= +2.8$,
$\mu_n= -1.9$. Theoretical estimates give the strength constant for
nucleon-nucleus weak potential $g_p \approx +4.5$ for a proton and
$|g_n|\sim 1$ for a neutron \cite{FlaMur97}. The aim of anapole
measurements is to provide  accurate values for these constants, thus obtaining important information about hadronic weak coupling.

A number of theoretical investigations of the nuclear spin-dependent parity violation in diatomic molecules have been performed in recent years, using both semiempirical \cite{KozFomDmi87,DmiKhaKoz92,KozLab95,DemCahMur08} and \textit{ab initio} methods \cite{TitMosEzh96,KozTitMos97,NayDas09,BakPetTit10,IsaHoeBer10,IsaBer12}  .
In a recent paper \cite{BorIliDzu12} we presented Dirac Hartree-Fock and relativistic density-functional calculations 
of the electronic $W_{A}$ factor of the diatomic group-2 and -12 fluorides and a number of 
other diatomic compounds. In this work we investigate the nuclear spin-dependent parity violation effects in a different type of system, i.e. positively charged dimers, 
as these systems have an experimental advantage of being easily trapped \cite{Odom11}. Diatomic ions have also been proposed for the search for the electron electric-dipole moment (eEDM) \cite{PetMosIsa07,MeyBoh08,IsaPetMos05}, and the preliminary experiments are currently being conducted \cite{CosGreSin12}.
Here we use the combination of methods presented in Ref. \cite{BorIliDzu12} to calculate the $W_{A}$ factors of positively 
ionized group-13 and group-15 fluorides ($^2\Sigma_{1/2}$ and $^2\Pi_{1/2}$ ground states, respectively), 
and a number of other positive diatomic ions having a $^2\Sigma_{1/2}$ ground state. 

%
For $^2\Sigma_{1/2}$ and $^2\Pi_{1/2}$ electronic states, the interaction (\ref{Ha}) can be replaced by the effective operator, 
which appears in the spin-rotational Hamiltonian \cite{FlaKri85_2,DemCahMur08},
\begin{eqnarray}
H_A^\mathrm{eff}=\kappa_{NSD} W_A\frac{(\mathbf{n}\times\mathbf{S}^\prime)\cdot \mathbf{I}}{I},
\label{eq:Heff}
\end{eqnarray}
where $\mathbf{S}^\prime$ is the effective spin and
$\mathbf{n}$ is the unit vector directed along the molecular axis  
from the heavier to the lighter nucleus. The electronic factor $W_A$ is found from evaluating the matrix
elements of the $\bm{\alpha}\rho(\mathbf{r})$ operator in the
molecular spinor basis \cite{Visscher1997181}. 
The $^2\Sigma_{1/2}$ and the $^2\Pi_{1/2}$ open-shell electronic states are
two-fold degenerate, corresponding to the two possible projections of
electronic angular momentum along $\mathbf{n}$,
i.e.~$|\Omega\rangle=|\pm\frac{1}{2}\rangle$.  
When operating within this degenerate space, the operator
$\frac{G_F}{\sqrt{2}}\bm{\alpha}\rho(\mathbf{r})$ is equivalent to
$W_A(\mathbf{n}\times\mathbf{S}^\prime$) (Eq.~(\ref{eq:Heff})).  
Time-reversal symmetry ensures that only the matrix elements that are
off-diagonal in $\Omega$ are non-vanishing. This symmetry rule is
encapsulated within the effective operator $H_A^\mathrm{eff}$ by the
angular factor $(\mathbf{n}\times\mathbf{S}^\prime)$. Here the
effective spin $\mathbf{S}^\prime$ generates rotations in the
degenerate subspace analogously to usual spin operator $\mathbf{S}$ in
a spin-1/2 system. 

The calculations
were carried out within the open-shell single determinant
average-of-configuration Dirac-Hartree-Fock approach (DHF)
\cite{Thyssen_thesis} and within the relativistic density functional theory (DFT) \cite{SauHel02}, employing quaternion  
symmetry \cite{Saue:1997, Saue:1999}. A finite nucleus, modeled by the Gaussian 
charge distribution was used \cite{VisDya97}. All the calculations were performed using the developer's version of the DIRAC10 program package \cite{DIRAC10}. 

For the lighter elements (boron to phosphorus), uncontracted aug-cc-pVTZ basis sets were
used \cite{KenDunHar92,WooDun93}. For the rest of the atoms, we employed
Faegri's dual family basis sets \cite{Fae01}. As a good description of the electronic
wave function in the nuclear region is essential for obtaining reliable
results for parity violating properties \cite{LaeSch99}, we
augmented the basis sets with high exponent $s$ and $p$ functions, which
brings about an increase of around $10\%$ in the calculated values of $W_{A}$.
The basis sets were increased, both in the core and in the valence regions, to
convergence with respect to the calculated $W_{A}$ constants. The final basis sets can be found in Table \ref{tab:I}.

\begin{table}
  \caption{Basis sets employed in the calculation of the $W_{A}$ constants. All
elements with $Z>15$ are described by the Faegri basis sets \cite{Fae01}
augmented by high exponent, diffuse, and high angular momentum functions.}
  \label{tab:I}
  \centering
\begin{tabular}
[c]{lrr}\toprule
Atom\ \ \  &\ \ \ \  $Z$ &\hspace{2.5cm} Basis Set\\\colrule
B & 5 & aug-cc-PVTZ\footnote{augmented by 3 high exponent $p$ functions}\\
N & 7 & aug-cc-PVTZ\footnote{augmented by 4 high exponent $s$ and 3 high exponent $p$ functions}\\
O & 8 & aug-cc-PVTZ\\
F & 9 & aug-cc-PVTZ\\
Al & 13 & aug-cc-PVTZ\footnote{augmented by 4 high exponent $p$ function.}\\
P & 15 & aug-cc-PVTZ\footnote{augmented by 1 high exponent $s$ and 3 high exponent $p$ functions}\\
Ga & 31 & 22\textit{s}19\textit{p}10\textit{d}8\textit{f}2\textit{g}\\
As & 33 & 21\textit{s}20\textit{p}11\textit{d}8\textit{f}2\textit{g}\\
Y & 39 & 21\textit{s}20\textit{p}12\textit{d}9\textit{f}2\textit{g}\\
Zr & 40 & 21\textit{s}20\textit{p}12\textit{d}9\textit{f}2\textit{g}\\
In & 49 & 22\textit{s}20\textit{p}12\textit{d}9\textit{f}2\textit{g}\\
Sb & 51 & 22\textit{s}21\textit{p}13\textit{d}9\textit{f}2\textit{g}\\
Hf & 72 & 25\textit{s}22\textit{p}16\textit{d}10\textit{f}2\textit{g}\\
Tl & 81 & 25\textit{s}23\textit{p}15\textit{d}10\textit{f}2\textit{g}\\
Bi & 83 & 25\textit{s}24\textit{p}16\textit{d}11\textit{f}2\textit{g}\\
Ac & 89 & 26\textit{s}24\textit{p}16\textit{d}11\textit{f}2\textit{g}\\
\hline\hline
\end{tabular}
\end{table}
Where available, experimentally determined bond distances $R_{e}$ were used. 
For molecules where $R_{e}$ is not known experimentally, we optimized the bond
distance using relativistic coupled cluster
theory with single, double, and perturbative triple excitations,
CCSD(T) \cite{Visscher:1996}. To reduce the computational effort,
we employed an infinite order two-component relativistic
Hamiltonian obtained after the Barysz--Sadlej--Snijders (BSS) transformation of
the Dirac Hamiltonian in a finite basis set \cite{IliJenKel05,Ilias:2007}. 
Our calculated $R_{e}$ are typically within 0.01 \AA \ of the experimental values, where available.
The experimental/calculated equilibrium distances can be found in Table \ref{tab:II}.

\begin{table}
 \caption{Internuclear distances $R_{e}$ (taken from CCSD(T) calculations, unless referenced otherwise, \AA), core-polarization scaling parameters $K_{CP}$, the $P$-odd interaction constants $W_{A}$ (Hz) obtained using DHF and DFT, and the final recommended values, taken as $W_A$(Final)$=(W_A$(DFT)$K_{CP}+W_A$(DHF)$K_{CP}$)/2. Relativistic factors $R_W$ (see Eq.~(\ref{Rw}))  are also shown.}
  \centering
  \begin{tabular}{llllllll}
    \hline\hline
     & $Z$ &$R_W$ &  $R_e$ (\AA )& $K_{CP}$ & \multicolumn{3}{c}{$W_A$ (Hz)}  \\
    \cline{6-8}
                   &               &           &         &     & DHF & DFT & Final                     \\
    \hline
    \\
    \multicolumn{8}{c}{Group 13 ($^{2}\Sigma_{1/2}$)} \\
    BF$^+$  &  5 & 1.01 &   1.314 &1.1& 1.74 & 1.71    & 1.90 \\
    AlF$^+$  &  13 & 1.07 &   1.590\footnotemark[1] & 1.2 & 9.62 & 10.39   &12.0     \\
    GaF$^+$  &  31 & 1.41 &   1.683\footnotemark[2] &1.1&94.4& 93.5& 103.4 \\
    InF$^+$  &  49 & 2.2  &   1.91 &1.1& 370.3&   358.3 & 400.7    \\
    TlF$^+$ &  81 & 7.4  &   2.00 &1.1& 3833& 3622   & 4100 \\
    \\
    \multicolumn{8}{c}{Group 15 ($^{2}\Pi_{1/2}$)} \\  
    NF$^+$  &  7 & 1.02  &    1.180\footnotemark[3]  &1.1 & --0.014& --0.015&--0.016\\ 
   PF$^+$  &  15  & 1.10 &   1.524 &1.2 & --0.16& --0.18  & --0.20  \\
    AsF$^+$ & 33 & 1.47   &   1.660 &1.1 & --6.48& --7.29  & --7.67    \\
    SbF$^+$ & 51& 2.3   &   1.832 &1.1 & --60.0& --66.1 & --71.6  \\
    BiF$^+$  & 83 & 8.1 &   2.281 &1.1& --2204& --2123 & --2380 \\   
    \\
    \multicolumn{8}{c}{Other systems ($^{2}\Sigma_{1/2}$)} \\  
     YF$^+$ &  39  & 1.69 &    1.885 &1.2 & 45.6& 39.4 & 51.0  \\ 
   ZrO$^+$ &  40& 1.73   &    1.695 &1.2 &43.2&31.1& 44.6  \\ 
    HfO$^+$ & 72 & 5.0   &   1.708 &1.2 & 662.5& 609.1 & 762.9 \\
    AcF$^+$ & 89 & 11    &   2.106 &1.2 &1654& 1614& 1961 \\  
    \hline\hline
  \end{tabular}
\footnotetext[1]{Ref.~\cite{DykKirMor84}}
\footnotetext[2]{Ref.~\cite{YosHir95}}
\footnotetext[3]{Ref.~\cite{Cha98}}
  \label{tab:II}
\end{table}
In the DFT calculations we used the Coulomb-attenuated B3LYP functional 
(CAMB3LYP*), the parameters
of which were adjusted by Thierfelder \textit{et al.} \cite{ThiRauSch10} to reproduce the PV energy shifts obtained
using coupled cluster calculations (the newly adjusted parameters are $\alpha= 0.20$,
$\beta=0.12$, and $\mu=0.90$). 

In our previous work \cite{BorIliDzu12} we have examined and compared various schemes for adding electron correlation to the 
Dirac--Hartree--Fock $W_A$ values, and core-polarization contributions to the DFT results. Here, we correct the calculated DHF and 
DFT $W_A$ for core polarization using a scaling parameter, $K_{CP}$. This parameter is obtained from atomic calculations as described 
in the following. 
The main contribution to the matrix elements of the NSD interaction for the 
valence molecular electrons comes from short distances around the heavy nucleus, where the total molecular potential is spherically symmetric to very high precision, and the core of the heavy atom is practically unaffected by the presence of the second atom, justifying our use of the atomic model. The molecular orbitals of the valence electron can thus be expanded in this region, using spherical harmonics centered at
the heavy nucleus,
\begin{equation}
  |\psi_v \rangle= a |s_{1/2} \rangle +b |p_{1/2} \rangle + c|p_{3/2} \rangle + d|d_{3/2} \rangle \dots
\label{eq:psi_v}
\end{equation}
Only $s_{1/2}$ and $p_{1/2}$ terms of this expansion give significant
contribution to the matrix elements of the weak interaction.
These functions can be considered as states of an atomic
 valence electron and are calculated
using standard atomic techniques in two different approximations: one
that includes electron correlation and another that does not.

The single electron DHF Hamiltonian is given by
\begin{equation}
  \hat H_0 = c \bm{\alpha} \cdot \mathbf{p} + (\bm{\beta} -1)c^2 -
  \frac{Z}{r} + V_e(r), 
\label{eq:h0}
\end{equation}
where $\bm{\alpha}$ and $\bm{\beta}$ are the Dirac matrices and $V_e(r)$ is the self-consistent DHF potential due to atomic electrons.

The self-consistent DHF procedure is first performed for the closed shell ion, from
which the valence electron is removed. Then the core potential $V_{\rm
  DHF}^{N-N_v}$ is frozen and the valence $s_{1/2}$ and $p_{1/2}$ states
are calculated by solving the DHF equation for the valence electron,
\begin{equation}
(\hat H_0 - \epsilon_v) \psi_v=0,
\label{eq:DHF}
\end{equation}
where $\hat H_0$ is given by (\ref{eq:h0}).

The core polarization can be understood as the change of the
self-consistent DHF potential due to the effect of the extra term (the
weak interaction operator $\hat H_{\rm A}$) in
the Hamiltonian. The inclusion of the core polarization in
a self-consistent way is equivalent to the random-phase
approximation (RPA, see, e.g.~\cite{DzuFlaSil87}). The change in the DHF
potential is found by solving the RPA-type equations self-consistently
for all states in the atomic core, 
\begin{equation}
(\hat H_0 - \epsilon_c) \delta \psi_c = -(\hat H_{\rm A} + \delta V_{\rm A})\psi_c.
\label{eq:RPA}
\end{equation}
Here, $\hat H_0$ is the DHF Hamiltonian (\ref{eq:h0}), index $c$
enumerates the states in the core, $\delta \psi_c$ is the correction to the
core state $c$ due to weak interaction $\hat H_{\rm A}$, and $\delta V_{\rm A}$
is the correction to the self-consistent core potential due to the
change of all core functions. Once $\delta V_{\rm A}$ is found, the core polarization can be
included into a matrix element for valence states $v$ and $w$ via the
redefinition of the weak interaction Hamiltonian,
\begin{equation}
\langle v |\hat H_{\rm A}| w \rangle \rightarrow \langle v |\hat H_{\rm
  A} + \delta V_{\rm A}| w \rangle. 
\label{eq:meRPA}
\end{equation}
We then obtain the scaling parameter for core-polarization effects, $K_{CP}$, from

\begin{equation}
  K_{CP} = \frac{\langle \psi^{\rm DHF}_{ns_{1/2}}  | \hat H_{\rm A} + \delta
    V_{\rm A}
    |\psi^{\rm DHF}_{n^{\prime}p_{1/2}} \rangle}{\langle \psi^{\rm DHF}_{ns_{1/2}}|
    \hat H_A |\psi^{\rm DHF}_{n^{\prime}p_{1/2}} \rangle} .
\label{eq:K_CP}
\end{equation}

It should be noted that for the positively charged group 15 fluorides we have only calculated the correlations between the valence electrons and the core; the correlations between the valence $ns$ and $np$ electrons are not included.


We investigated two types of positively ionized diatomic molecules: those with a $^{2}\Sigma_{1/2}$ ground state, including group 13 fluorides and a number of other systems, and molecular ions with a $^{2}\Pi_{1/2}$ ground state (represented here by group 15 fluorides). The values of $K_{CP}$ for all the systems under study are presented in Table \ref{tab:II}, together with the DHF and the DFT $W_A$ constants. 
As the final recommended value for the $W_A$ parameter we take an average of $W_A$(DHF)$K_{CP}$ and $W_A$(DFT)$K_{CP}$. The estimate of the accuracy in our previous work \cite{BorIliDzu12} has shown that it is about 15\% for molecules in the $^2\Sigma_{1/2}$ electronic state and 20-30\% for the $^2\Pi_{1/2}$ state.

The magnitude of $W_{A}$ in the $^2\Sigma_{1/2}$ electronic state is expected to scale as $Z^{2}R_W$ \cite{FlaKri85}, where $R_W$ is the relativistic parameter,
\begin{align}
R_W  & =\frac{2\gamma+1}{3}\left(  \frac{a_{B}}{2Zr_{0}A^{1/3}}\right)
^{2-2\gamma}\frac{4}{\left[  \Gamma(2\gamma+1)\right]  ^{2}},\label{Rw}\\
\gamma & =[1-(Z\alpha)^{2}]^{1/2}.\nonumber
\end{align}
\begin{figure}
  \centering
  \includegraphics[width=1.0\linewidth]{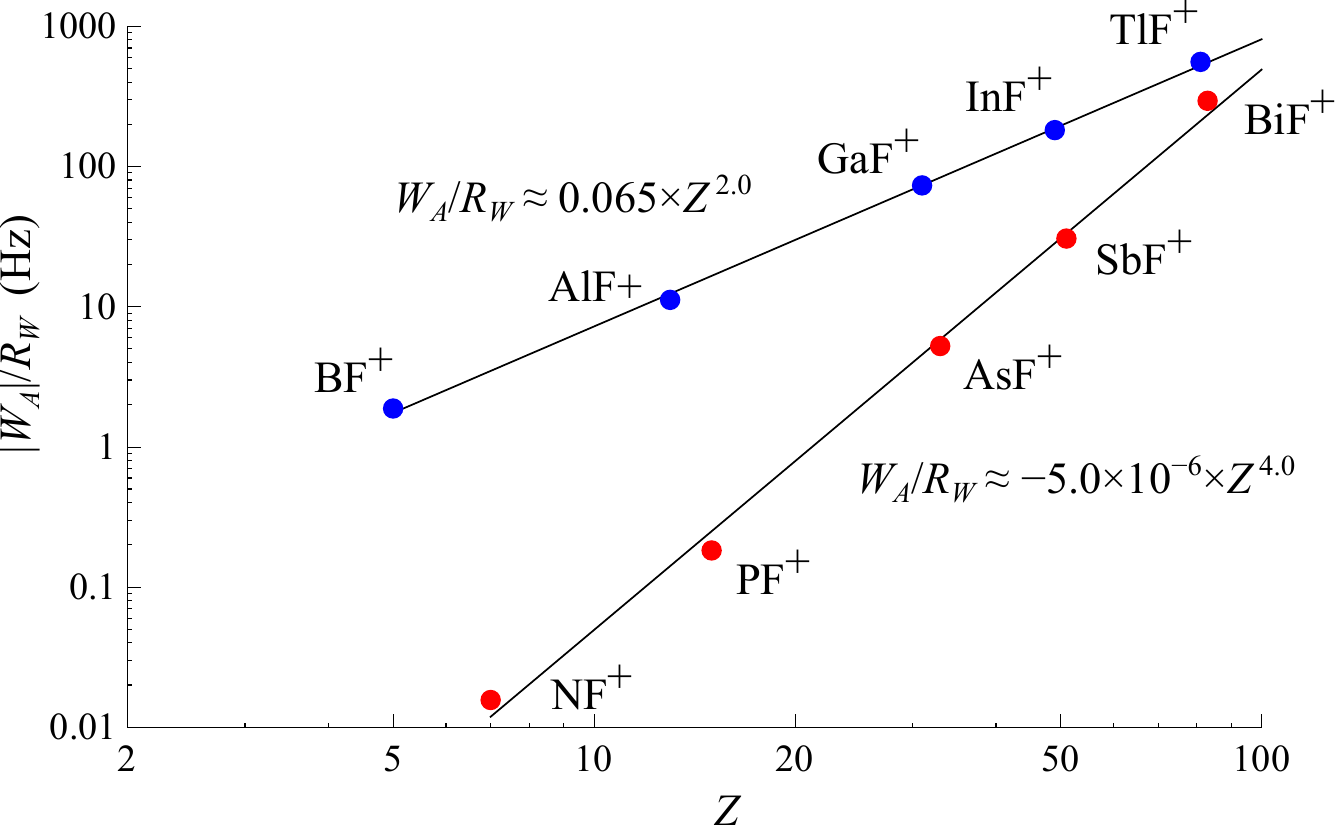}
  \caption{ (color online) Log-log plot illustrating the scaling of $W_A$ with
relativistic parameter $R_W$ and atomic number $Z$ for group-13 and -15
singly-ionized fluorides.}
  \label{fig:I}
\end{figure}
In Eq.~(\ref{Rw}), $a_{B}$ is the Bohr radius, $r_{0}=1.2\times10^{-15}$
m, and $\alpha$ is the
fine-structure constant. The $R_{W}$ parameters are shown in Table
\ref{tab:II} for each of the metal atoms. In Fig. \ref{fig:I} we plot $\log\left(
\frac{|W_{A}|}{R_{W}}\right)  $ as a function of $\log(Z)$ for both groups of
dimers. For group-13 fluorides the scaling is, indeed, $Z^{2}$. 
In the case of group 15 fluorides, however, the ground state is $^{2}\Pi_{1/2}$, for which the $W_A$ parameter vanishes in the non-relativistic limit, since in this limit it does not contain the $s$-wave electronic orbital and can not provide the matrix element $\langle s_{1/2}|\bm{\alpha}\rho(\mathbf{r})|p_{1/2}\rangle$. The effect appears due to the mixing of  $^2\Sigma_{1/2}$ and $^2\Pi_{1/2}$ electronic states by the spin-orbit interaction, and gives an extra factor of $Z^2 \alpha^2$ in the $Z$-dependence of $W_A$, as seen in Fig \ref{fig:I}.  


To summarize, here we have performed calculations of the $P$-odd interaction
constant $W_A$ in singly-ionized group-13 and group-15 fluorides as well as other
select singly-ionized diatomic systems. To the best of our knowledge, this is the
first investigation of nuclear spin-dependent parity violation effects in molecular
ions and it is motivated by progress in the cooling and trapping of such systems.


This work was supported by the Marsden Fund (Royal Society of New Zealand), the Australian Research Council,
the Alexander von Humboldt Foundation (Bonn), and the Slovak Research and Development Agency (grant number APVV-0059-10). The authors are grateful to R. Berger and T. Isaev for critical comments.

%


\end{document}